\documentclass[prl,twocolumn,showpacs,preprintnumbers,amsmath,amssymb]{revtex4-1} 
\usepackage{dcolumn}
\usepackage{bm}

\usepackage{amsmath}
\usepackage{amssymb}
\usepackage{amsfonts} 
\usepackage{wasysym}  
\usepackage{graphics}  
\usepackage{psfrag} 
\usepackage{graphicx} 
\usepackage{epsfig}    
\usepackage{rotating}  
 \usepackage[latin1]{inputenc}
 \usepackage{color}
 \usepackage{rotating}
\usepackage{csquotes}
 \usepackage{multirow}
\usepackage[nooneline, tight,raggedright,FIGTOPCAP]{subfigure}
\usepackage[linkcolor=red,citecolor=blue,urlcolor=blue,colorlinks=true]{hyperref}
\usepackage{color}
\usepackage{epsfig}
\usepackage{bibunits}
\usepackage{flushend}
\usepackage{balance}

\definecolor{darkgreen}{rgb}{0,0.5,0} 
\definecolor{violet}{rgb}{0.5,0,0.5}
\definecolor{orange}{rgb}{0.2,0.5,0.5}
\newcommand{\avg}[1]{\langle #1\rangle}

\newcommand{\bequ}{\begin{equation}}
\newcommand{\eequ}{\end{equation}}
\newcommand{\bequa}{\begin{eqnarray}}
\newcommand{\eequa}{\end{eqnarray}}
\newcommand{\bse}{\begin{subequations}}
\newcommand{\ese}{\end{subequations}}
\newcommand{\tn}[1]{\textnormal{#1}}

\begin{document}
\begin{bibunit}
\title{A nonequilibrium diffusion and capture mechanism ensures tip-localization of regulating proteins on dynamic filaments}
\author{Emanuel Reithmann}

\author{Louis Reese}
\altaffiliation{Present address:  Department of Bionanoscience, Delft University of Technology, Delft, Netherlands.}
\author{Erwin Frey}
\email{frey@lmu.de}

\affiliation{
Arnold Sommerfeld Center for Theoretical Physics (ASC) and Center for NanoScience (CeNS), Department of Physics, Ludwig-Maximilians-Universit\"at M\"unchen, Theresienstrasse 37, 80333 M\"unchen, Germany
}

\begin{abstract}
Diffusive motion of regulatory enzymes on biopolymers with eventual capture at a reaction site is a common feature in cell biology. Using a lattice gas model we study the impact of diffusion and capture for a microtubule polymerase and a depolymerase. Our results show that the capture mechanism localizes the proteins and creates large-scale spatial correlations. We develop an analytic approximation that globally accounts for relevant correlations and yields results that are in excellent agreement with experimental data. Our results show that diffusion and capture operates most efficiently at cellular enzyme concentrations which points to \emph{in vivo} relevance.
\end{abstract}

\date{\today}

\maketitle

The diffusive motion of proteins on filamentous structures in the cell is vital for several cellular functions such as gene regulation~\cite{Hippel1989} and cytoskeletal dynamics~\cite{Cooper2009,Howard2007}:
To find their target sites, transcription factors are likely to employ one-dimensional diffusion on the DNA and the dynamics of this process largely determine the kinetics of gene regulation~\cite{Riggs1970,Hammar2012}. 
Similarly, actin and microtubule (MT) binding proteins diffuse on the respective filaments and fulfill regulatory functions primarily at the filament ends. 
Adam and Delbr\"uck~\cite{Adam1968} suggested that a reduction in dimensionality of the diffusive motion enhances the effective rate of association of particles with binding sites on the membrane or on DNA and filaments, and this concept has been widely applied and extended~\cite{Richter1974,*Berg1981,Bintu2005}, see also Refs.~\cite{Mirny2009,*Kolomeisky2011,*Benichou2011,*Sheinman2012}  for recent reviews on the topic.

With regard to cytoskeletal architectures, efficient association and localization of enzymes to specific sites is relevant for a variety of cellular processes throughout the cell cycle and for cell motility and dynamics~\cite{Dos2003,*Akhmanova2008}.
It was recently shown experimentally that one-dimensional diffusion is utilized~\cite{Helenius2006,Cooper2009} by two proteins with important roles in the regulation of MT dynamics~\cite{Tournebize2000,Kinoshita2001,Wilbur2013,Reber2013}, MCAK and XMAP215. These proteins strongly localize at their respective reaction sites and show association rates for these sites that are significantly higher than expected for binding via three-dimensional diffusion~\cite{Helenius2006,Brouhard2008}. Both  proteins carry out vital tasks, with MCAK acting as depolymerase of tubulin protofilaments~\cite{Desai1999} and XMAP215 as a poylmerase~\cite{Brouhard2008} when bound to ends of MTs. Note that similar mechanisms are also assumed to be relevant for actin associated proteins~\cite{Romero2004,*Vavylonis2006,*Hansen2010,*Mizuno2011}. 
However, diffusive motion on filaments does not lead to a localization and efficient association of proteins \emph{per se}: 
As we have shown previously~\cite{Reithmann2015}, it is crucial to include a \emph{capturing mechanism} at the reaction site, which suppresses the one-dimensional diffusive motion of a protein that reaches this site; without such a capturing mechanism no increase in the effective association rate for the tip occurs. For MCAK and XMAP215 protein capturing is observed in experiments: Diffusive motion stops once the proteins reach the MT tip~\cite{Brouhard2008,Helenius2006}. Yet, the underlying interactions with the MT tip are still elusive and being studied~\cite{Friel2011,*Ritter2016}.

Here we present a theoretical description of enzyme diffusion and capture at MT tips where the enzymes catalyze filament polymerization or depolymerization. 
Previous studies of similar systems have lacked either a capturing mechanism~\cite{Klein2005,Schmitt2011} or a dynamic filament~\cite{Reithmann2015}, although both features are critical.
To overcome both limitations, we employ a one-dimensional lattice gas~\cite{Chou2011, Derrida1998} with particle capturing in a dynamic system, in which growth or shrinkage of the filament is triggered by the interactions of particles with the lattice end. 
Our motivation is twofold: Firstly, we seek for a detailed mathematical understanding of the capturing mechanism. 
Secondly, based on a fully quantitative model, we wish to elucidate the specific biomolecular mechanisms employed by XMAP215 and MCAK.
Our results show that the capturing process induces large-scale spatial correlations in the protein distribution along the filament. We develop a mathematical framework that systematically includes relevant correlations on a global scale. This conceptual advancement allows us to quantitatively explain the results of \emph{in vitro} experiments with XMAP215 and MCAK~\cite{Brouhard2008,Cooper2010}. We demonstrate that the diffusion and capture mechanism strongly localizes XMAP215 and MCAK at the MT tip and that the process operates optimally under physiological conditions for both proteins, which suggests that it is relevant \emph{in vivo}.

\begin{figure}[!b]
\centering
\includegraphics[width = \columnwidth]{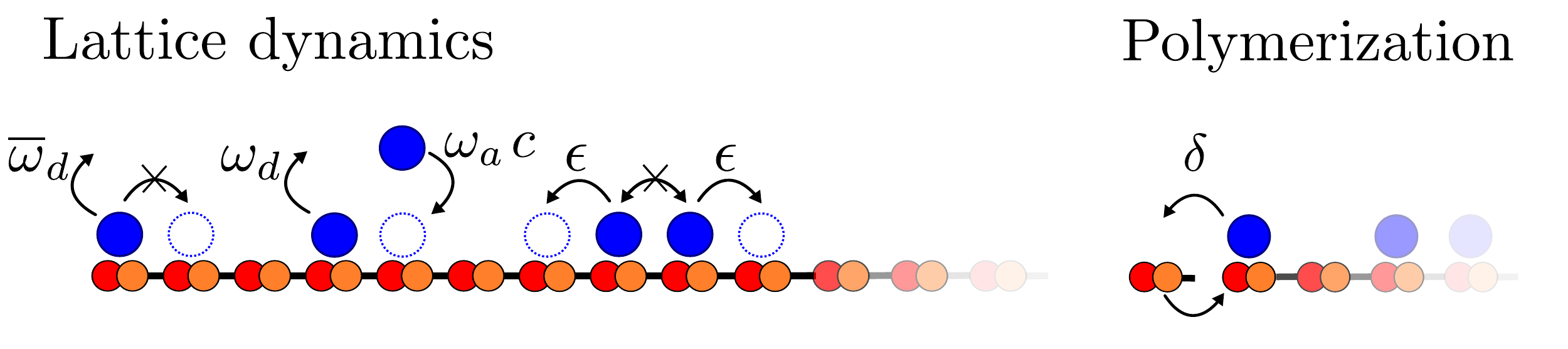}
\caption{\label{fig:model}\label{fig:cartoon}
Illustration of the model for XMAP215. Particles bind to empty lattice sites with rate $\omega_\tn{a} c$, where $c$ is the particle concentration in solution, and detach with rate $\omega_d$. The proteins hop symmetrically to neighboring sites at rate $\epsilon$ but exclude each other. We assume a distinct off-rate $\overline{\omega}_\tn{d}$ at the first site. Particles bound there cease hopping but add new lattice sites at rate $\delta$. The particle which stimulates polymerization moves with the tip. An analogous model can be defined for MCAK, where depolymerization occurs if the lattice end is occupied, see Supporting Material for details~\cite{SuppMat}.
}
\end{figure}

\paragraph{Model definition.} We consider a one-dimensional lattice with lattice spacing $a$ and a semi-infinite geometry which corresponds to one protofilament, as depicted in Fig.~\ref{fig:model}. In the case of MTs, $a$ is the length of a tubulin heterodimer, $8.4\,\tn{nm}$.
The configuration of enzymes on the lattice is described by occupation numbers $n_i$, taking values $n_i{=}0$ for empty, and $n_i{=}1$ for occupied sites. 
The particles symmetrically hop to neighboring sites at rate $\epsilon$, and interact via hard-core repulsion. 
We implement Langmuir kinetics to model a surrounding reservoir of particles with a constant concentration $c$. Particles attach to and detach from the lattice at rates $\omega_\tn{a} c$ and $\omega_\tn{d}$, respectively~\cite{Lipowsky2001,Parmeggiani2003,*Parmeggiani2004}. 
Sites $i {\ge} 3$ are considered as \emph{bulk} sites. There the dynamics differs from that in the bulk as we implement a protein capturing mechanism: Hopping from site $i{=}1$ to site $i{=}2$ is disallowed, as suggested experimentally for MCAK and XMAP215~\cite{Helenius2006,Brouhard2008}. In this way, detailed balance is broken which leads to strong tip-localization due to a particle flux along the filament; in equilibrium models such a significant localization is absent, see Fig.~\ref{fig:TipLocalozation} in the Supporting Material~\cite{SuppMat}.
Particles detach from the first lattice site at a distinct off-rate, $\overline{\omega}_\tn{d} {\neq} \omega_\tn{d}$.
We refer to site $i {=} 1$ as a \emph{reaction} site at which new lattice sites may be added or removed.
For the moment, we specify our discussion to polymerases such as XMAP215~\cite{Brouhard2008}. However, our considerations are largely independent of whether polymerization or depolymerization occurs---an equivalent formulation can also be found for the depolymerase MCAK~\cite{SuppMat}. 
For XMAP215, we specify that lattice growth is triggered at rate $\delta$ if the protein is bound to the first lattice site. 
Hence, the average speed of lattice growth $v$ for the MT is proportional to the average particle occupation $\avg{n_1}$ and the XMAP215 polymerization rate: $v {=} \delta a \avg{n_1}$. Here we assume one catalyzing protein per protofilament end at saturating conditions~\cite{SuppMat}. The actual maximum number of catalytically active proteins is unknown;  in experimental literature approximately 10 XMAP215 proteins at the MT tip are estimated at $50 \, \tn{nM}$ XMAP215~\cite{Brouhard2008}.
As shown in recent experiments, XMAP215 acts \emph{processively}, i.e. one molecule adds multiple tubulin dimers to the growing MT end~\cite{Brouhard2008}. 
To implement such behavior in our model the particle at the tip is transferred to newly incorporated lattice sites. 
In our analysis we neglect uncatalyzed tubulin addition or removal as typical corresponding experiments~\cite{Helenius2006, Cooper2010, Brouhard2008, Widlund2011} were performed under conditions where these processes did not occur with a significant rate. An extension is, however, possible in a straightforward fashion and does not affect tip-localization significantly; see Fig.~\ref{fig:SpontGrowth} and~\ref{fig:DensityDynamicLattice} in the Supporting Material~\cite{SuppMat}. Therefore we expect validity of our further considerations also with intrinsic MT dynamics, for example as a consequence of hydrolysis of tubulin bound GTP which was studied extensively in previous models~\cite{Antal2007,*Padinhateeri2012,*Niedermayer2015}. 

\paragraph{Mathematical analysis.} We set up the equations of motion for the average occupation numbers of the stochastic process defined above. All equations will be formulated in the frame of reference comoving with the dynamic lattice end. In the bulk of the lattice, $i {\geq} 3$, we obtain
\bequa
\label{eq:Xbasis1}
\tfrac{\tn{d}}{\tn{d} t}{\avg{ {n}_i }}&=& \epsilon  \bigl(\avg{n_{i+1}} {-} 2 \avg{n_i} {+} \avg{n_{i-1}} \bigr) 
+ \delta \bigl( \avg{ n_1 n_{i-1}} {-} \avg{ n_1 n_{i} } \bigr) 
\nonumber \\
&&+\, \omega _\tn{a} c \bigl( 1 {-} \avg{n_i} \bigr) 
- \omega _\tn{d} \avg{ n_i } \, .
\eequa
This equation comprises contributions from hopping while obeying the exclusion principle~\cite{Derrida2007} (terms proportional to $\epsilon$) and a displacement current due to polymerization (terms proportional to $\delta$) as well as particle attachment and detachment (terms proportional to $\omega_a$ and $\omega_d$, respectively). The tip occupations complement these bulk dynamics in the following manner:
\begin{eqnarray}
\label{eq:XMAP}
\tfrac{\tn{d}}{\tn{d} t}\rho_1
&=& \epsilon (\rho_2{-}g_2) + \omega _\tn{a} c (1 {-} \rho_1) - \overline{\omega} _\tn{d} \rho_1 \, ,  \\
\tfrac{\tn{d}}{\tn{d} t} \rho_2 
&=& \epsilon (\rho_3 {-} 2 \rho_2 {+} g_2) -  \delta g_2 \nonumber +  \omega _\tn{a} c ( 1 {-} \rho_2) - \omega_\tn{d} \rho_2 \, , \nonumber \\
\tfrac{\tn{d}}{\tn{d} t} g_2
&=& \epsilon  (g_3 {-} g_2) + \delta g_2 + \omega_\tn{a} c (\rho_1 {+} \rho_2 {-} 2 g_2 ) - (\omega_\tn{d} {+} \overline{\omega}_\tn{d}) g_2 \, . \nonumber 
\end{eqnarray}
Here we have defined the average density, $\rho_i {:=} \langle n_i \rangle$, and the correlation function, $g_i {:=} \langle n_1 n_i \rangle$. Moreover, since the polymerization process facilitated by XMAP215 is processive, an empty lattice site at $i{=}2$ is created and site $i{=}1$ remains occupied each time a new site is added.
We fully quantify our model with the experimental data available for XMAP215~\cite{Brouhard2008,Widlund2011}; see Supporting Material for parameter values~\cite{SuppMat}.

\begin{figure}[htb]
\centering
\includegraphics[width = \columnwidth]{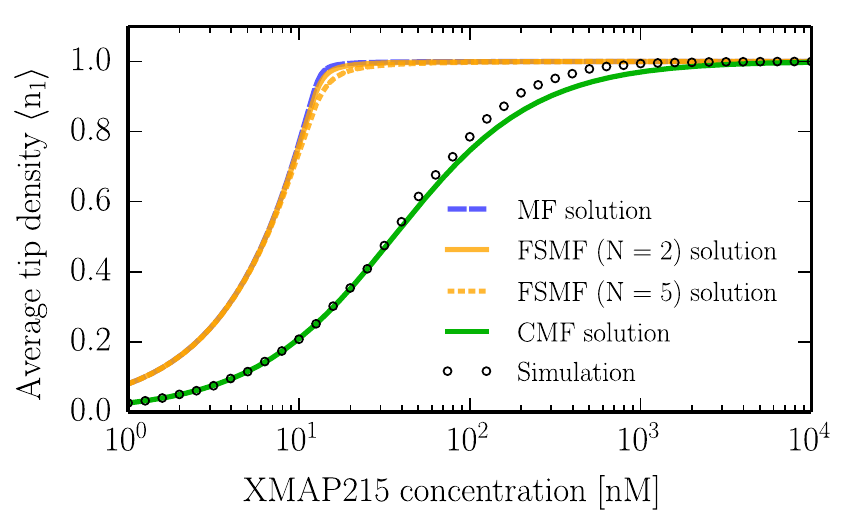}
\caption{
Average occupation of the first lattice site $\avg{n_1}$. The MF approach as well as the FSMF approximation for segment sizes of $N{=}2,5$ deviate strongly from stochastic simulation data (open circles), in complete contrast to the CMF approximation. Parameter values are detailed in the Supporting Material~\cite{SuppMat}.
}
\label{fig:mf_sols}
\end{figure}

In the first step we test the quality of standard approximation techniques for driven lattice gases against stochastic  simulation data obtained from Gillespie's algorithm~\cite{Gillespie2007}.
The set of equations which determines the lattice occupations (Eq.~\ref{eq:Xbasis1} and Eqs.~\ref{eq:XMAP}) is not closed; the dynamics of the density $\rho_i$ and the correlation functions $g_i {=} \langle n_1 n_i \rangle$ are coupled. In fact, there is a hierarchy of equations, which, in general, precludes the derivation of an exact solution for many driven lattice gas systems. 
A common and often quite successful approximation scheme for exclusion processes is to assume that there are no correlations and that one may factorize all correlation functions, $\langle n_1 n_i \rangle {\approx} \langle n_1 \rangle  \langle n_i \rangle$. In this \emph{mean-field} (MF) approximation one obtains a closed set of differential equations for the particle density $\rho_i$ which may be solved subject to proper boundary conditions; see Supporting Material for details. Fig.~\ref{fig:mf_sols}  shows the average occupation number of the first site, $\langle n_1 \rangle$, as a function of the protein concentration in solution $c$. A comparison with our stochastic simulation data shows that the MF solution strongly overestimates $\langle n_1 \rangle$ and thus the average polymerization speed $v$. 

One possible reason for the failure of the MF calculation lies in correlations that arise close to the reaction site. Local correlations can efficiently be accounted for by employing a \emph{finite segment mean-field} (FSMF) theory~\cite{Chou2004,Lakatos2005}. Here, the idea is to retain all correlations close to the catalytic site by solving the full master equation for the first $N$ sites and to use the MF assumption only outside of this segment. The density profile is then obtained by matching the tip solution and the MF solution~\cite{Klein2005, Nowak2007}; see Supporting Material~\cite{SuppMat}. While the results show the right trend towards the numerical data, the improvement over the MF results is insignificant. These observations suggest that correlations extend far beyond the immediate vicinity of the reaction site.

To account for such correlations we extend the MF theory by retaining both the density and the correlation function as dynamic variables. In order to close the set of equations we employ the following factorization scheme: $\avg{n_1 n_2 n_i} {\approx} \avg{n_1 n_2} \avg{n_i}$, and $\avg{n_2 n_i} {\approx} \avg{n_2} \avg{n_i}$ for $i {\geq} 3$, i.e. we retain correlations with respect to the reaction site but neglect them within the bulk of the lattice. We confirmed this approximation scheme for typical biological parameter values by stochastic simulations; see Fig.~\ref{fig:MomentFactorization} in the Supporting Material~\cite{SuppMat}. With the above closure relations one obtains for the bulk dynamics in a continuous description
\bse\label{eq:CMFPDE}\begin{eqnarray}
\partial_t \rho (x,t) 
&=& D \partial_x^2 \rho 
  - v_0 \partial_x g 
  + \omega _\text{a} c (1 {-}  \rho)  
  - \omega _\text{d} \rho \, ,
 \label{eq:rhobulk}\\
\partial_t g (x,t) 
&=& D \partial_x^2 g - v_0 \partial_x g 
    +\epsilon \rho \, \bigl(\rho_2 {-} g_2 \bigr) 
\nonumber \\
&& + \, \omega_\tn{a} c \, \bigl( \rho {+} \rho_1 {-} 2 g \bigr)
   - \bigl( \omega_\tn{d} {+} \overline{\omega}_\tn{d} \bigr) g \, , 
\label{eq:Cx}
\end{eqnarray}\ese
where we defined $\rho(x,t) {=} \avg{n_{i+1}}$ and $g(x,t) {=}\avg{n_1 n_{i+1}}$ with $x{=}a(i{-}1)$ for $i {\geq} 3$. We have further introduced the macroscopic diffusion constant $D {=}\epsilon  a^2$ and the maximum polymerization speed $v_0 {=} \delta  a$. Equations~\ref{eq:CMFPDE} can be derived from the discrete equations for the density $\rho_i$, Eq.~\ref{eq:Xbasis1}, and the correlation function $g_i$; for details see the Supporting Material~\cite{SuppMat}. Due to the capturing mechanism a continuous description is not valid at sites $i{=}1,2$, and we retain the local dynamics there, Eqs.~\ref{eq:XMAP}. These equations constrain the boundary conditions of $\rho(x)$ and $g(x)$ at $x{=}a$. We further impose that the density equilibrates asymptotically at the Langmuir isotherm, $\lim_{x \to \infty}\rho(x) {=} \rho_\tn{La} {=} \omega_\tn{a} c/ (\omega_\tn{a} c {+} \omega_\tn{d})$, and that correlations vanish, $\lim_{x \to \infty}g(x) {=} \avg{n_1} \rho_\tn{La}$.
Solving the equations of this \emph{correlated MF} (CMF) theory for the steady state tip density we obtain the results shown in Fig.~\ref{fig:mf_sols}, which are in excellent agreement with the stochastic simulation data. We therefore conclude that there are long-ranged correlations along the MT and that they are essential in explaining the observed average tip density and the ensuing polymerization speed. 

\begin{figure}[htb]
\centering
\includegraphics[width = \columnwidth]{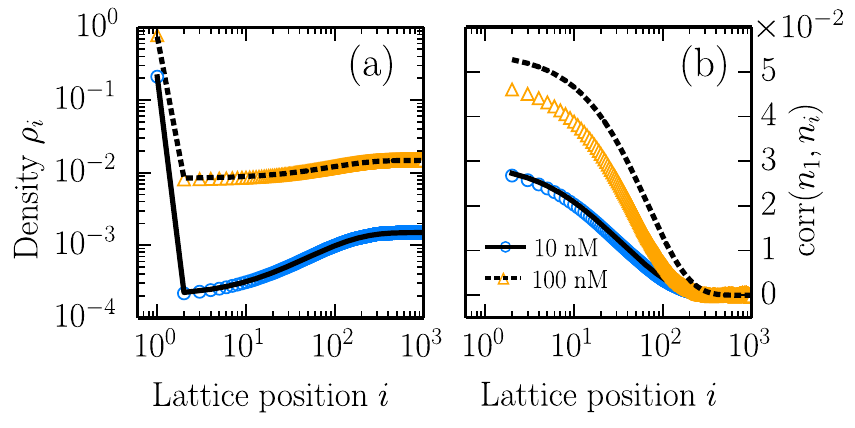}
\caption{
Comparison of density and tip-bulk correlation profiles obtained by the CMF approximation (lines) and stochastic simulations (symbols) for XMAP215 concentrations of $10$ and $100\,\text{nM}$ (see Supporting Material~\cite{SuppMat} for parameter values~\cite{Brouhard2008, Widlund2011}). (a) XMAP215 strongly localizes to the MT tip and the density profile drops abruptly at sites $i{=}1,2$. (b) The correlation coefficient $\mathrm{corr}(n_1, n_i)$ (see Eq.~\protect{\eqref{eq:corr}}) along the lattice shows the significance of tip-bulk correlations over hundreds of lattice sites.
\label{fig:XMAP_correl}\label{fig:XMAP_density}}
\end{figure}

Fig.~\ref{fig:XMAP_density}(a) shows the density profile along the lattice  obtained by stochastic simulations and the CMF approach. The particle occupation is obtained with high precision within the CMF framework along the whole lattice. The density profiles also agree with recent data from time and ensemble averaged high resolution fluorescence intensity profiles for XMAP215~\cite{Maurer2014}. Notably, there is a discontinuity at sites $i{=}1,2$, which is due to particle capture and which demonstrates the strong tip-localization of the proteins. 

\begin{figure}[t]
\centering
\includegraphics[width = \columnwidth]{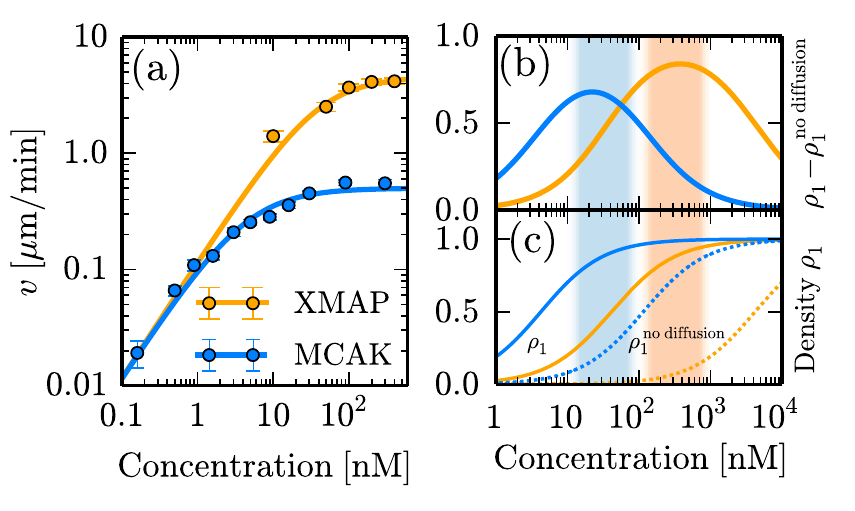}
\caption{Panel (a) demonstrates excellent agreement of polymerization and depolymerization velocities obtained from our theoretical analysis (CMF approximation) with existing experimental data for XMAP215~\cite{Brouhard2008,Widlund2011} and MCAK~\cite{Cooper2010}, respectively. 
Panel (b) depicts the difference between the occupation density at the tip $\rho_1$  with and without diffusion on the MT, where $\rho_1^\tn{no\ diffusion} = \omega_\tn{a} c/(\overline{\omega}_\tn{d}+\omega_\tn{a} c)$, and shows the impact of diffusion and capture on tip localization of MCAK (blue) and XMAP (orange). The concentration range for maximum efficiency coincides with the physiological concentration range for each protein: $100-1000\,\tn{nM}$ for XMAP215~\cite{Kinoshita2001} and $10-100\,\tn{nM}$ for MCAK~\cite{Hunter2003} (shaded areas). In (c) the reaction site density with lattice diffusion ($\rho_1$, solid lines) and without lattice diffusion ($\rho_1^\tn{no\ diffusion}$, dashed lines) is depicted.
Kinetic parameters are given in the Supporting Material~\cite{SuppMat}.\label{fig:EXP}}
\end{figure}
In Figure~\ref{fig:XMAP_correl}(b), the Pearson product-momentum correlation coefficient 
\bequa
\tn{corr} (n_1, n_i)=\frac{\tn{cov}(n_1, n_i)}{\sigma(n_1) \sigma(n_i)},
\label{eq:corr}\eequa
which quantifies the correlations between the tip site $i{=}1$ and sites $i{\geq}2$ in the bulk, is plotted against lattice position. Here $\tn{cov} (\cdot,\cdot)$, and $\sigma(\cdot)$ signify the covariance and the standard deviation, respectively.
The correlation coefficient decays very slowly over a broad region at the tip.
The capturing mechanism and the resulting particle flux towards the filament tip ensue strong positive correlations with respect to the first lattice site and sites in its vicinity. 
This effect is antagonized by weak negative correlations caused by the creation of empty lattice sites due to polymerization. 
With diffusion taking place on a faster time scale than polymerization, the positive correlations dominate. 
This is confirmed by stochastic simulations where either capturing or growth is switched off: We find anti-correlations if capturing is turned off, and positive correlations if there is no growth of the lattice; see Fig.~\ref{fig:CorrelBehavior} in the Supporting Material~\cite{SuppMat}. Note that for higher growth rates the correlation profile can also become negative.
We conclude that the spatial correlations which emerge over several hundred lattice sites are a direct consequence of protein capture and processive growth. 
Further, it becomes evident why the MF and the FSMF approaches do not lead to the correct tip density:
Correlations extend into the system on a length-scale which exceeds the scope of these and other previous approaches~\cite{Schmitt2011,Klein2005,Reithmann2015}. In contrast the CMF approximation captures and quantifies significant correlations and successfully reproduces simulation data.  Note that also higher order correlations of the form $\avg{n_1 n_j n_k}$ and $\avg{n_j n_k}$ with $j,k{\geq}2$ and $k{>}j$, which are neglected in the CMF approximation, might be of relevance when particle interactions become important for lattice diffusion. This explains the deviations in the computed correlation profile, Fig~\ref{fig:XMAP_correl}(b). As the CMF method is based on a non-perturbative ansatz there is no analytic expression that exactly quantifies its error. However, we observe very good agreement with our Gillespie algorithm based simulations over a very broad parameter range and, importantly, for typical biological parameters, see Fig.~\ref{fig:CMFDev} in the Supporting Material~\cite{SuppMat}.

\paragraph{Comparison with experimental data.}

We now turn to a comparison with experimental data for the polymerization velocity~\cite{Brouhard2008,Widlund2011} and, to supplement the results for XMAP215, we apply our methods to an analogous model for MCAK particles which depolymerize MTs~\cite{Helenius2006,Cooper2010}. 
In essence, we adapt the above model to account for lattice shrinkage triggered by an occupied reaction site, see Supporting Material for details~\cite{SuppMat}. Similar to the processive polymerization of XMAP215 also MCAK is assumed to depolymerize processively~\cite{Helenius2006, Cooper2010}. The parameters employed in the model are again drawn from available experimental data~\cite{Cooper2010}.
For both MCAK and XMAP215, we find excellent quantitative agreement between our theoretical approach and experimentally determined polymerization and depolymerization velocities; see Fig.~\ref{fig:EXP}(a). This quantitative agreement is achieved without an adjustable parameter; see Supporting Material~\cite{SuppMat}.
We then used the quantified models to investigate the impact of the diffusion and capture process for XMAP215 and MCAK. Fig.~\ref{fig:EXP}(b) shows the increase of protein localization at the reaction site due to diffusion and capture on the filament: 
We plot the difference between tip densities in the presence ($\rho_1^\tn{CMF}$) and absence of diffusion on filaments ($\rho_1^\tn{no\ diffusion} {=} \omega_\tn{a} c/(\overline{\omega}_\tn{d}{+}\omega_\tn{a} c)$). 
For both enzymes, diffusive motion and subsequent capturing at the MT lattice strongly increases the occupation density at the tip and therefore constitutes a highly efficient means of increasing the effective attachment rate to the reaction site.
Moreover, the ensuing curve shows a pronounced maximum, indicating an optimal concentration range at which the enhancement of tip occupancy due to diffusion on the MT reaches its peak.
Strikingly, this maximum coincides with the physiological concentration range for each protein: $100{-}1000\,\tn{nM}$ for XMAP215~\cite{Kinoshita2001} and $10{-}100\,\tn{nM}$~\cite{Hunter2003} for MCAK. This strongly supports the importance of diffusion and capture for MCAK and XMAP215 \emph{in vivo}. 

It is interesting to speculate about possible biomolecular mechanisms that could generate particle capturing at the MT tip as such a mechanism would probably require an energy source to drive the system out of equilibrium.
Concerning MCAK, it was recently hypothesized, that an ATP is required to stop its diffusive motion at the MT tip~\cite{Friel2011,Ritter2016} which is consistent with our proposed non-equilibrium model. Since XMAP215 does not bind nucleotides such as ATP or GTP itself~\cite{Brouhard2008}, one might speculate that a non-equilibrium capturing mechanism relies on tubulin polymerization or depolymerization.
Possibly, a conformational change of XMAP215 coupled to processes involved in MT depolymerization or polymerization could lead to protein capture.

\paragraph{Summary and Conclusion.} 
In this work, we studied the regulatory influence of an explicit capture process on the distribution of MT polymerases and depolymerases that are subject to one-dimensional diffusion on MTs. 
To model these biologically relevant situations we employed a model based on a \emph{symmetric simple exclusion process}~\cite{Derrida1998} extended by a detailed balance breaking capturing process at the lattice end, which acts as a biasing mechanism.
Our results show that the occupation of the MT tip with a protein spatially correlates with the occupation of the MT lattice. This is a direct consequence of protein capturing which in turn strongly localizes the proteins at the MT tip.
Correlations decay slowly along the lattice and have a large impact on the occupation of the MT tip. This is of relevance as the latter quantity determines the velocity of enzyme-dependent MT growth or shrinking. 
We derive a generalized set of hydrodynamic equations which couple the evolution of the particle density with the evolution of relevant correlations.
In that way it is possible to account for those correlations on a global scale. Similar correlations have been identified in two-dimensional diffusive systems~\cite{Markham2013a,*Markham2013b} or in diffusive systems with a small, local drive~\cite{Sadhu2014}. 

Our findings are not limited to MTs and their associated enzymes, but might also be applicable to other enzymatic processes with spatial degrees of freedom and, quite generally, non-equilibrium physics.
 
\begin{acknowledgments}
We thank Linda Wordeman, Gary Brouhard and their respective co-workers for sharing data points of XMAP215 and MCAK induced MT growing and shrinking velocities, respectively, as shown in Fig.~\ref{fig:EXP}(a).
This research was supported by the German Excellence Initiative via the program ``NanoSystems Initiative Munich'' (NIM) and the Deutsche Forschungsgemeinschaft (DFG) via project B02 within the Collaborative Research Center (SFB 863) ``Forces in Biomolecular Systems''.
\end{acknowledgments}

\end{bibunit}
\clearpage
\balance
\begin{bibunit}
\onecolumngrid
\begin{center}
\textbf{\large Supplemental Material: A nonequilibrium diffusion and capture mechanism ensures tip-localization of regulating proteins on dynamic filaments}\\
\vspace{0.3cm}
Emanuel Reithmann, Louis Reese, and Erwin Frey\\
\textit{Arnold Sommerfeld Center for Theoretical Physics (ASC) and Center for NanoScience (CeNS), Department of Physics, Ludwig-Maximilians-Universit\"at M\"unchen, Theresienstrasse 37, 80333 M\"unchen, Germany}
\end{center}
\onecolumngrid

\setcounter{equation}{0}
\setcounter{figure}{0}
\setcounter{table}{0}
\setcounter{page}{1}
\makeatletter
\renewcommand{\theequation}{S\arabic{equation}}
\renewcommand{\thefigure}{S\arabic{figure}}
\renewcommand{\bibnumfmt}[1]{[R#1]}
\renewcommand{\citenumfont}[1]{R#1}
\section{Tip localization due to particle capturing}
\begin{figure}[!ht]
\centering
\includegraphics[width = 0.65\columnwidth]{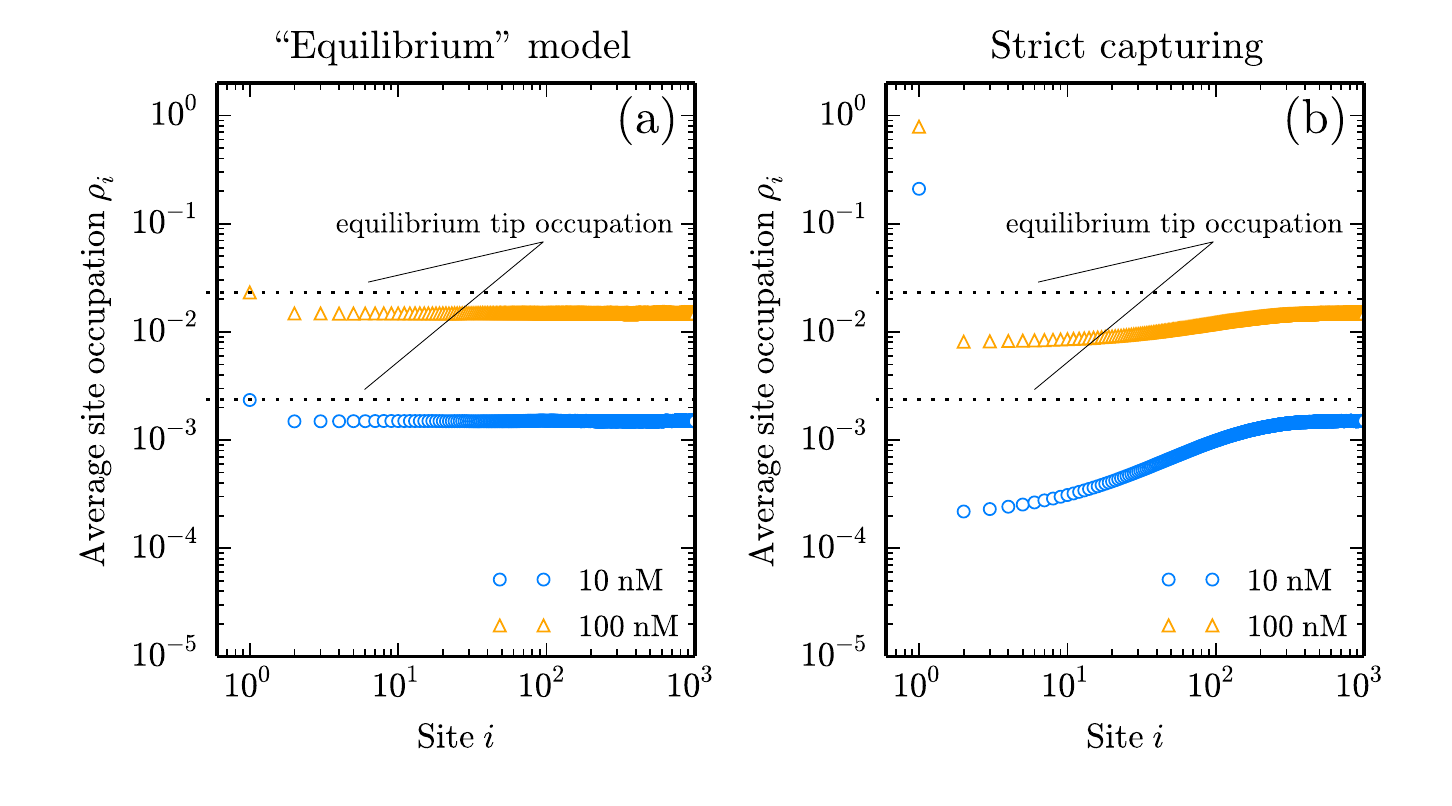}
\caption{Diffusion and capture ensures tip-localization. Density profiles from MC simulations (open symbols) of (a) a model where particle hopping obeys detailed balance with respect to a static lattice and (b) the model from the main text. In an ``equilibrium" model no localization occurs and the density profile is almost constant due to fast diffusion. With strict (i.e. irreversible capturing), the tip is highly occupied as compared to its equilibrium occupation (dotted lines). Note that in both models we implement off-rates which differ at the tip and lattice growth which results in non-constant density profiles also when particle hopping obeys detailed balance. Further, also the  ``equilibrium" model is out of equilibrium due to lattice growth. In (a) $\overline{\epsilon} = 3.0 \times 10^{3}\, \mathrm{s}^{-1}$, in (b) $\overline{\epsilon} = 0$. Other parameters as for the XMAP215 model, see Table~\ref{tab:MCAK_rates}. \label{fig:TipLocalozation}}
\end{figure}

In this work, we investigate a model where the diffusive motion of particles on a filament ceases as soon as they arrive at a reaction site. This feature, which we refer to as particle capturing, is a key element of our model, as it drives the system out of thermal equilibrium. In order to investigate the impact of particle capture on tip localization of particles, we also investigated a model where particles are not captured at the tip, but where a hopping from the tip into the bulk occurs such that detailed balance is not broken. In detail, we introduce a release rate $\overline{\epsilon}$, at which particles hop from site $i=1$ to site $i=2$. Then, to implement equilibrium conditions for particle hopping (i.e. with respect to a system without lattice growth or shrinkage), we impose $\epsilon/\overline{\epsilon} = \omega_d/\overline{\omega}_d$. This condition ensures detailed balance in a static system and for a constant on-rate along the lattice. Since we also implement lattice growth, detailed balance is still broken, which manifests itself in a net particle drift away from the tip in the comoving frame of reference. In Fig.~\ref{fig:TipLocalozation} we compare density profiles of the hopping-equilibrium model and the one with strict (i.e. irreversible) particle capturing as defined in the main text with parameters as for XMAP215. In the equilibrium model, the density profile is almost constant whereas in the model with capturing a strong tip-localization occurs (1-2 orders of magnitude increase in the tip-density). Although an irreversible capturing is, of course, a simplification, we expect similar effects to occur for release rates much smaller than the equilibrium release rate, $\overline{\epsilon} \ll \overline{\epsilon}_\mathrm{eq}:=(\epsilon\, \overline{\omega}_d)/\omega_d$. In this case, capturing generates a particle current towards the MT tip which conversely leads to spatial correlations subject of this work.

\section{Mean-field (MF) approximation}
In the mean-field approximation all correlations are neglected; we set $\avg{n_{i} n_{j}} = \avg{n_i} \avg{n_j}$. This closes the hierarchy of equations stated in the main text:
\begin{eqnarray}
	 \tfrac{\tn{d}}{\tn{d} t}{\avg{ {n}_i }}&=& \epsilon (\avg{n_{i+1}} -2 \avg{n_i} + \avg{n_{i-1}}) + \delta (\avg{ n_1} \avg{ n_{i-1}} - \avg{ n_1} \avg{ n_{i} })+ \omega _\tn{a} c ( 1 - \avg{n_i}) - \omega _\tn{d} \avg{ n_i }  \ \mathrm{for\ } i \geq 3 \label{eqn:sup_ni}\\	 
 \tfrac{\tn{d}}{\tn{d} t}{\avg{ {n}_1 } }&=& \epsilon (\avg{n_2}-\avg{n_1 } \avg{n_2}) + \omega _\tn{a} c (1 - \avg{n_1}) - \overline{\omega} _\tn{d} \avg{ n_1}  \label{eqn:sup_n1} \\
 \tfrac{\tn{d}}{\tn{d} t}{\avg{ {n}_2}} &=& \epsilon (\avg{ n_3} - 2 \avg{n_2} + \avg{n_1} \avg{ n_2 }) -  \delta \avg{ n_1 } \avg{n_2 } +  \omega _\tn{a} c ( 1 - \avg{n_2}) - \omega _\tn{d} \avg{n_2} \, . \label{eqn:sup_n2} 
\end{eqnarray}
Instead of solving the recurrence relation, we use a continuous description for Eq.~\ref{eqn:sup_ni}. At sites $i=1,2$ such an approximation is not valid due to a discontinuity in the density profile. Performing a Taylor expansion for small lattice spacings $a$ up to second order we obtain
\bequa
 \partial_t \rho (x,t) &=&  \epsilon a^2 \partial_x^2 \rho (x,t)  - \delta a \partial_x \rho(x,t) \avg{n_1} + \omega _\tn{a} c (1 -  \rho (x,t) )  - \omega _\tn{d} \rho(x,t)\, .
\eequa
In the above equation the continuous labeling $x = a (i-1)$ is used for $\rho(x,t) = \avg{n_{i+1}} $. Further, we use that for typical biological systems $\epsilon \gg \delta$ holds true and neglect the second order term due to the particle drift in the comoving frame, $\tfrac{1}{2} \delta a^2 \partial^2_x \rho(x)$. Since we are only interested in the steady state solution we set the time derivative to zero. As boundary condition, we impose that the density equilibrates at the Langmuir density for large distances to the tip, $\lim_{x \to \infty}\rho(x) = \rho_\mathrm{La} = \omega_a c / (\omega_a c + \omega_d)$. The boundary condition at $x=a$ has to be consistent with the solution of Eqs.~\ref{eqn:sup_n1} and~\ref{eqn:sup_n2},  $\rho(a) = \avg{n_2}$. We can use the continuous solution to express $\avg{n_3} = \rho(2a)$ and solve Eqs.~\ref{eqn:sup_n1} and~\ref{eqn:sup_n2}. This self-consistent solution can be obtained numerically and determines the MF density profile along the whole lattice. 
\section{The finite segment mean-field (FSMF) approximation}
The finite segment mean-field approach is based on the idea to account for correlations locally within a small segment. In detail, all correlations within this segment are retained whereas outside the segment correlations are neglected. An efficient implementation is achieved by using the transition matrix corresponding to the master equation for occupations of the segment. Since in our model correlations are strongest close to the tip, we choose to keep correlations with respect to the first $N$ sites. For example, for $N=2$ the corresponding transition matrix $M_{ij}$ with $i,j \in \{0,\dots,3\}$ reads
\bequa
M&=&\left(
 \begin{matrix} 
 -2 \omega _a c - \epsilon \avg{n_3} & \omega_d + \epsilon \avg{\overline{n}_3} & \overline{\omega}_d & 0 \\
 \omega _a c + \epsilon \avg{n_3} & - \omega_d -\omega_a c  - \epsilon (1 + \avg{\overline{n}_3} )  & 0 & \overline{\omega}_d \\
 \omega_a c & \epsilon & -\overline{\omega}_d -\omega_a c  -\epsilon \avg{n_3} & \delta + \epsilon \avg{\overline{n}_3} + \omega_d \\
 0 & \omega_a c & \omega_a c  + \epsilon \avg{n_3} & - \omega_d-\overline{\omega}_d  - \epsilon \avg{\overline{n}_3}  - \delta \\
 \end{matrix}  \right) .\nonumber  
\eequa
Here we introduced $\avg{\overline{n}_3} = (1- \avg{n_3})$. Further, the enumeration of states is chosen such that it corresponds to the respective binary number, e.g. $M_{01}$ describes transitions from state $(n_1=0, n_2=1)$ to state $(n_1=0, n_2=0)$. Note that correlations with respect to $n_{N+1}$ are already neglected. The eigenvector of $M$ with eigenvalue 0 is then computed, which yields steady state occupations within the segment in dependence of $\avg{n_{N+1}}$. A self-consistent solution of these occupations and those for sites $i > N$ is obtained in analogous fashion to the MF procedure: We use the continuous MF solution for densities with $i > N$ and the discrete solutions for sites in the segment to express all densities in terms of $\avg{n_{N+1}}$. The master equation for $\avg{n_{N+1}}$ (given by Eq.~\ref{eqn:sup_ni}) is then solved numerically in the steady state to compute the complete density profile. This procedure is, however, strongly limited by the size of the finite segment as the corresponding transition matrix is of size $2^{N}\times2^{N}$.
\section{The correlated mean-field (CMF) approximation}
In the following we will show how to perform the CMF approximation for the model presented in the main text. This approach systematically includes the relevant correlations arising due to the capturing mechanism. 

The CMF calculations can be separated in three steps: a) Computation of the continuous solution for the density $\rho(x)$ and correlation profile $g(x)$ in the bulk, $i\geq2$. b) Computation of the discrete solution for $i=1$. c) Matching of the continuous solution and the discrete solution.

\begin{figure}[!ht]
\centering
\includegraphics[width = 0.7\columnwidth]{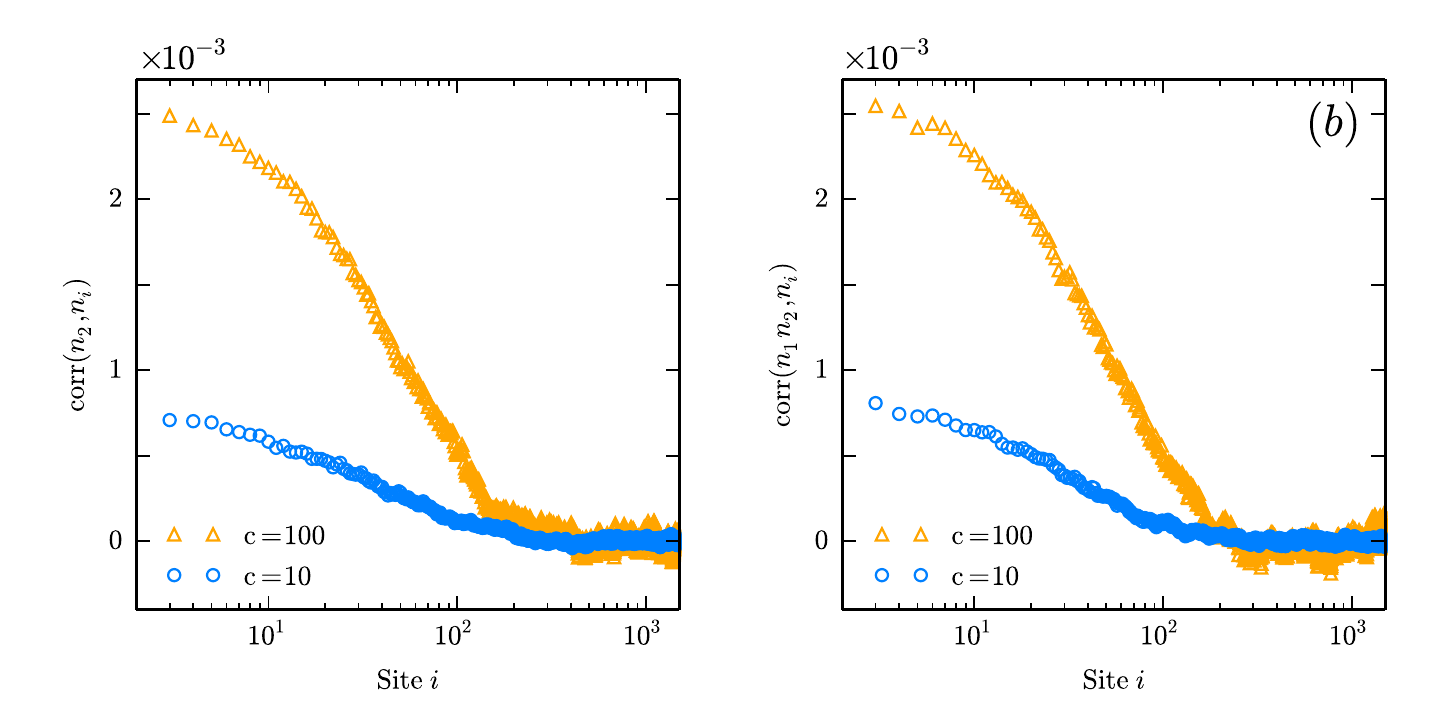}
\caption{In panel (a) and (b) we show that correlations $\mathrm{corr}(n_2,n_i)$ and $\mathrm{corr}(n_1 n_2, n_i)$ for $i \geq 3$ are negligible since they are one to two orders of magnitude smaller than the tip bulk correlations, $\mathrm{corr}(n_1,n_i)$. Parameters as for the XMAP215 model.\label{fig:MomentFactorization}}
\end{figure}

We start with deriving the continuous bulk solutions. The density profile is governed by Eq.~1 of the main text: 
\bequa
\label{eq:ni_correlated}
\tfrac{\tn{d}}{\tn{d} t}{\avg{ {n}_i }}&=& \epsilon  \bigl(\avg{n_{i+1} (1-n_i)} {-}  \avg{n_i (1-n_{i+1})} {+} \avg{n_{i-1}(1-n_i)} -\avg{n_{i}(1-n_{i-1})}\bigr) 
+ \delta \bigl( \avg{ n_1 n_{i-1}} {-} \avg{ n_1 n_{i} } \bigr) \nonumber \\
&&+\, \omega _\tn{a} c \bigl( 1 {-} \avg{n_i} \bigr) - \omega _\tn{d} \avg{ n_i } \nonumber \\
&=& \epsilon  \bigl(\avg{n_{i+1}} {-} 2 \avg{n_i} {+} \avg{n_{i-1}} \bigr) 
+ \delta \bigl( \avg{ n_1 n_{i-1}} {-} \avg{ n_1 n_{i} } \bigr) +\, \omega _\tn{a} c \bigl( 1 {-} \avg{n_i} \bigr) - \omega _\tn{d} \avg{ n_i } \, .
\eequa
Here, we account for particle hopping with exclusion (terms $\propto \epsilon$), lattice growth  (terms  $\propto \delta$), particle attachment  (terms  $\propto \omega_\tn{a}$), and particle detachment  (terms  $\propto \omega_\tn{d}$).
In the main text we show that it is essential to account for tip-bulk correlations on a large scale. In the CMF approach this is achieved globally by coupling the evolution of the density with the one for tip-bulk correlations. The discrete equation governing the evolution of correlations with respect to the reaction site reads 
\bequa
 \tfrac{\tn{d}}{\tn{d} t}{\avg{ n_1 n_i }}&=& \epsilon ( \avg{n_1 n_{i-1}} - 2 \avg{n_1 n_i} + \avg{n_1 n_{i+1}} + \avg{n_2 n_i} -\avg{n_1 n_2 n_i} )  + \delta ( \avg{n_1 n_{i-1}} - \avg{n_1 n_i} ) \nonumber \\
 && \, + \omega _\tn{a} c (\avg{n_1} + \avg{n_i} - \avg{n_1 n_i}) - ( \overline{\omega} _\tn{d} + \omega_\tn{d}) \avg{n_1 n_i}  . 
\eequa
The above equation, which follows from the master equation, describes changes of the joint probability for a simultaneous occupation of the first and the $i$-th site: All probabilities for processes that lead to a simultaneous occupation of both lattice sites multiplied with the respective rate are added and all probabilities for processes where one of the two sites is emptied multiplied with the respective rate are subtracted. Again, contributions arise from particle hopping with exclusion (terms  $\propto \epsilon$), lattice growth  (terms $\propto \delta$), particle attachment  (terms $\propto \omega_\tn{a}$), and particle detachment  (terms $\propto \omega_\tn{d}$), respectively. For example, for particle hopping we have contributions from hopping processes with respect to the $i$-th site ($\avg{n_1 n_{i-1}} - 2 \avg{n_1 n_i} + \avg{n_1 n_{i+1}}$) as well as the capturing of a particle at the first site ($\avg{n_2 n_i} -\avg{n_1 n_2 n_i} $). Note that higher order correlators can be obtained in complete analogy.
In order to close the hierarchy of moments, we use the factorization scheme stated in the main text: $\avg{n_1 n_2 n_i} \approx \avg{n_1 n_2} \avg{n_i}$ and $\avg{n_2 n_i} \approx \avg{n_2} \avg{n_i}$ for $i \geq 3$. Fig.~\ref{fig:MomentFactorization} shows that this is justified, as the corresponding correlation coefficients are one to two orders of magnitude lower than $\mathrm{corr}(n_1,n_i)$.
In the continuous limit $a\to 0$ the recurrence relations given by the dynamic equations for $\avg{n_i}$ and $\avg{n_1 n_i}$ translate into a set of coupled differential equations. Up to a second order Taylor expansion we obtain
\bequa
 \partial_t \rho (x,t) &=&  \epsilon a^2 \partial_x^2 \rho (x,t)  - \delta a \partial_x g(x,t) +  \omega _\tn{a} c (1 -  \rho (x,t) )  - \omega _\tn{d} \rho(x,t)\,  \label{eq:rhobulkSup}\\
 \partial_t g (x,t) &=& \epsilon ( a^2 \partial_x^2 g(x,t)+ \avg{n_2}(t) \rho(x,t) -\avg{n_1 n_2}(t)\rho(x,t) ) - \delta a \partial_x g(x,t)  + \omega_\tn{a} c (\avg{n_1}(t)+ \rho(x,t) - 2 g(x,t) )  \nonumber \\
 &&- (\omega_\tn{d} + \overline{\omega}_\tn{d}) g(x,t)\,.\label{eq:CxSup}
\eequa
Here, we used again a continuous labeling $x = a (i-1)$ and neglected second order terms due to lattice growth ($\propto \tfrac{1}{2} \delta a^2 \partial^2_x g(x)$) since $\epsilon \gg \delta$ for typical biological situations. In this work, we are interested in the steady state properties of the system, $\partial_t \rho(x,t)=0$ and $\partial_t g(x,t)=0$. Under this condition, Eqs.~\ref{eq:rhobulkSup} and~\ref{eq:CxSup} are solved for the continuous solutions $\rho(x)$ and $g(x)$. Further, we impose the following boundary conditions to obtain a meaningful solution: $\lim_{x\to \infty} \rho(x) = \rho_\tn{La}=\omega_\tn{a} c/ (\omega_\tn{a} c + \omega_\tn{d})$, $\lim_{x\to\infty} g(x) = \avg{n_1} \rho_\tn{La}$, $\rho(a)=\avg{n_2}$ and $g(a) = \avg{n_1 n_2}$. Note that the solutions depend on the yet unknown variables $\avg{n_1}$, $\avg{n_2}$ and $\avg{n_1 n_2}$. 

In the second step, we solve the equation for the occupancy of the reaction sites, $i=1$,
\bequa
   \label{eq:n1Sup}
 \tfrac{\tn{d}}{\tn{d} t}{\avg{ {n}_1 } }&=& 0 = \epsilon (\avg{n_2}-\avg{n_1 n_2}) + \omega _\tn{a} c (1-\avg{ n_1}) - \overline{\omega}_\tn{d} \avg{ n_1} \, ,
\eequa
to express $\avg{n_1}$ in terms of $\avg{n_2}$ and $\avg{n_1 n_2}$.

Lastly, we self-consistently match the discrete and continuous solutions in that we determine the values of $\avg{n_2}$ and $\avg{n_1 n_2}$. To this end we employ the ``master equations" for the latter variables. 
\bequa
 \label{eq:n2Sup}
 \tfrac{\tn{d}}{\tn{d} t}{\avg{ {n}_2}} &=& 0 =\epsilon (\avg{ n_3} - 2 \avg{n_2} + \avg{n_1 n_2}) -  \delta \avg{ n_1 n_2 } +  \omega _\tn{a} c (1- \avg{n_2 }) - \omega _\tn{d} \avg{n_2} \\
  \label{eq:n1n2Sup}
 \tfrac{\tn{d}}{\tn{d} t}{\avg{{n}_1 {n}_2 }}&=& 0 =\epsilon  (\avg{n_1 n_3} - \avg{n_1 n_2}) + \delta \avg{n_1 n_2} + \omega_\tn{a} c (\avg{n_1} + \avg{n_2} - 2 \avg{n_1 n_2} ) - (\omega_\tn{d} + \overline{\omega}_\tn{d}) \avg{n_1 n_2} .
\eequa
We insert the continuous bulk solutions derived in the first step for $\avg{n_3} = \rho(2a)$ and $\avg{n_1 n_3} = g(2 a)$. Finally, the discrete solution for $\avg{n_1}$ is used to express all variables in terms of $\avg{n_2}$ and $\avg{n_1 n_2}$. This allows us to solve Eqs.~\ref{eq:n2Sup} and~\ref{eq:n1n2Sup} numerically which, as a consequence, fixes the entire density and correlation profile.

\begin{figure}[!ht]
\centering
\includegraphics[width = 0.65\columnwidth]{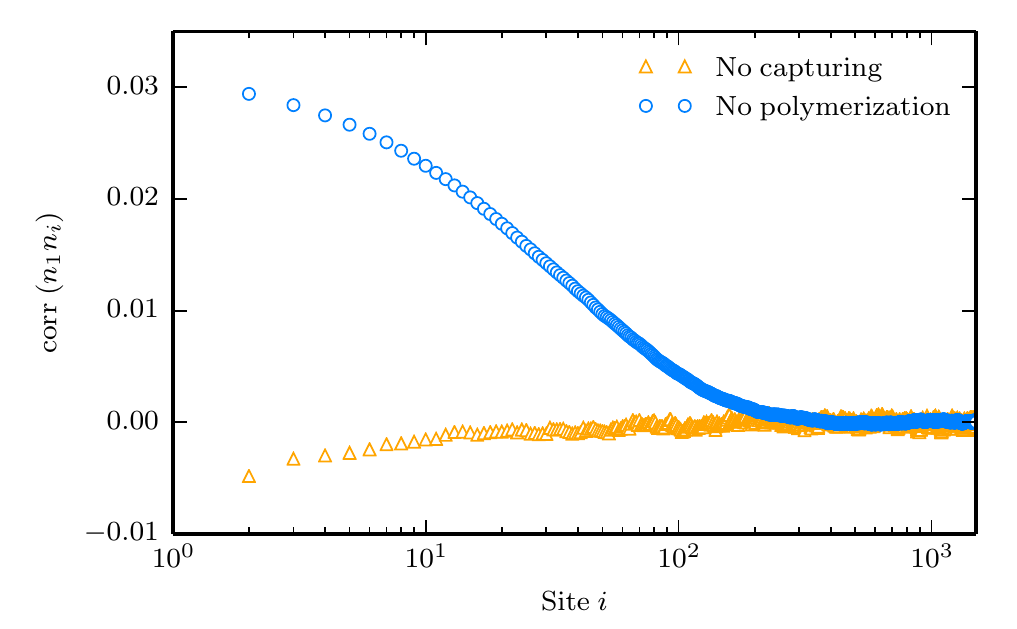}
\caption{Tip-bulk correlation profile obtained from stochastic simulations. Without particle capturing (orange data points) correlations are negative due to the processive growth of the lattice and the resulting creation of empty lattice sites. Correlations are positive in a static system with a capturing mechanism (blue points). Parameter values are equal to the ones used for the XMAP215 model; concentrations are $c=10\ \mathrm{nM}$ for the case without polymerization and $c=5000\ \mathrm{nM}$ for the case without capturing.\label{fig:CorrelBehavior}}
\end{figure}

The behavior of correlations is also demonstrated in Fig~\ref{fig:CorrelBehavior}: Without a capturing mechanism, correlations are purely negative due to the creation of empty sites resulting from the processive polymerization scheme. Opposed to that, purely positive correlations arise in a static lattice with capturing.

\begin{figure}[!ht]
\centering
\includegraphics[width = 0.65\columnwidth]{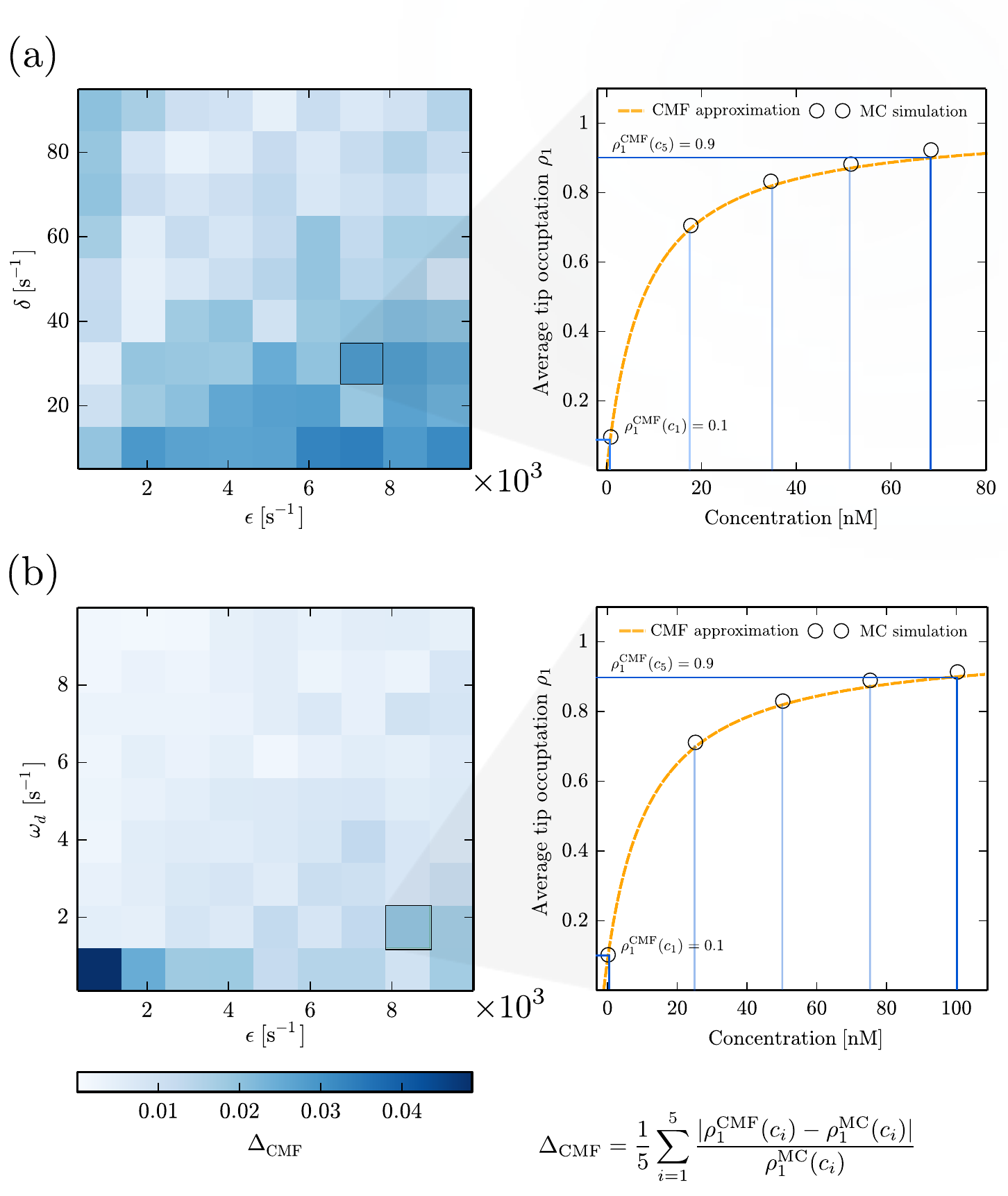}
\caption{Error of CMF approximation. We compared results for the tip density obtained from the CMF approximation ($\rho_1^\mathrm{CMF}$) and MC simulations ($\rho_1^\mathrm{MC}$) for 1000 different parameter sets. For each set $\{\epsilon$, $\delta$, $\omega_a$, $\omega_d$, $\overline{\omega}_d\}$ we determined five equidistant concentrations between $c_1$ and $c_5$, such that $\rho_1^\mathrm{CMF}(c_1) = 0.1$ and $\rho_1^\mathrm{CMF}(c_5) = 0.9$. For these concentrations, we computed the average relative deviation between simulation results and analytic approximation to get an estimate $\Delta_\mathrm{CMF}$ of the error along a $\rho_1-c$ curve (right side). Note that we expect the error to vanish for very low and very high occupations. We performed sweeps with respect to $\epsilon$ and $\delta$ (a), and $\epsilon$ and $\omega_d$ (b). Deviations are small, with the maximal $c$-averaged deviation being 5\% and the maximal relative deviation being 6.5\%. Color encodes the $c$-averaged deviations $\Delta_\mathrm{CMF}$ with white denoting $0$\% deviation and dark blue denoting more significant deviations. As expected, we observe a small trend of increasing errors whenever interactions in the lattice bulk become more frequent, i.e. for high $\epsilon$, small $\delta$ and small $\omega_d$. Opposed to Eq.~\ref{eq:CxSup} we include the second order term that arises due to lattice polymerziation, $\tfrac{1}{2} \delta a^2 \partial^2_xg(x)$,  as $\epsilon{\gg}\delta$ does not necessarily hold true any more.\label{fig:CMFDev}}
\end{figure}

The CMF approach neglects correlations within the diffusive compartment (i.e. we assume $\avg{n_i n_j} = \avg{n_i}\avg{n_j}$ and $\avg{n_1 n_i n_j} = \avg{n_1 n_i} \avg{n_j}$ for $i, j \geq 3$ and $i < j$). As this approximation is a non-perturbative ansatz, it is in general not possible to quantify its error. In order to ensure the validity over a broad and biologically relevant parameter range, we performed extensive MC simulations and compared the result with CMF computations. In detail, we performed parameter sweeps for $\epsilon$ (from $300-10000\ \mathrm{s}^{-1}$), $\omega_d$ (from $0.1-10\ \mathrm{s}^{-1}$), $\delta$ (from $5-95\ \mathrm{s}^{-1}$) and $c$ (for each parameter point at five equidistant values between $c_1$ and $c_5$, such that $\rho_1^\mathrm{CMF}(c_1) = 0.1$ and $\rho_1^\mathrm{CMF}(c_5) = 0.9$). The results are shown in Fig.~\ref{fig:CMFDev}. The CMF approximation delivers good results over this very broad parameter range; the maximum relative deviation for $\rho_1$ over the 1000 different tested parameter sets is 6.5\%.

In a previous publication, we derived an effective theory that allows for the calculation of reaction site occupations that are subject to a diffusion and capture mechanism in a static lattice (i.e. without lattice growth or shrinkage)~\cite{Reithmann2015}. While both approaches consider protein diffusion and capture on filaments, they differ significantly on a conceptual level and with respect to the scope of their predictions:
Whereas the previous approach is based an on a heuristic theory and \emph{a priori} only valid in the absence of polymerization and depolymerization, respectively, the CMF approach specifically accounts for lattice growth and shrinkage. Further, the CMF approximation is derived from more conceptual considerations: It assumes that diffusion and capture creates correlations which primarily affect the tip occupation while the diffusive motion of proteins on the MT depends less significantly on mutual correlations~\cite{Derrida2007}. As a consequence, the CMF approach yields density and tip-bulk correlation profiles for protein occupations along the MT, which are beyond the scope of our previous approach. As shown in the main text, the latter quantities are key to a quantitative understanding of tip-localization due to diffusion and capture and related processes.

\section{Uncatalyzed growth and shrinkage of MTs}

The model described in the main text does not account for MT growth or shrinkage in the absence of depolymerziation or polymerization factors like MCAK or XMAP215. The reason for this assumption is twofold: a) In the experiments with XMAP215~\cite{Widlund2011} and MCAK~\cite{Cooper2010} low concentrations of free tubulin were used such that no spontaneous MT growth was observed. Also, the measurements in Widlund et al.~\cite{Widlund2011} suggest that the rate of tubulin detachment in the corresponding experiments is negligible. b) Concerning MT depolymerization, we aim for a description of protein induced tubulin removal from stabilized MTs in analogy to \emph{in vitro} experiments with MCAK~\cite{Cooper2010,Helenius2006}. In this way, our model neglects the dynamic instability seen for unstabilized MTs~\cite{Antal2007,Padinhateeri2012,Niedermayer2015,Zakharov2015}, but provides a description how a stabilizing structure at the MT tip (e.g. GTP-tubulin) can be removed by regulatory enzymes.

That being said, let us emphasize that an extension towards uncatalyzed tubulin attachment and detachment is feasible based on the model described in the main text. To this end we include further processes in the model: If the terminal lattice site is unoccupied, a new site can be added at rate $\delta_\mathrm{spont}^\mathrm{poly}$ or removed at rate $\delta_\mathrm{spont}^\mathrm{depoly}$. For completeness, we also include catalyzed (processive) growth \emph{and} shrinkage with corresponding rates $\delta_\mathrm{cat}^\mathrm{poly}$ and $\delta_\mathrm{cat}^\mathrm{depoly}$, respectively. The resulting equations for the CMF framework then read
\bequa
 \partial_t \rho (x,t) &=&  0 = (\delta_\mathrm{spont}^\mathrm{depoly} -\delta_\mathrm{spont}^\mathrm{poly}) a \partial_x \rho(x,t) +  (\epsilon + \frac{1}{2} \delta_\mathrm{spont}^\mathrm{depoly} + \frac{1}{2} \delta_\mathrm{spont}^\mathrm{poly}) a^2 \partial_x^2 \rho (x,t)  \nonumber \\
 &&+( \delta_\mathrm{cat}^\mathrm{depoly} - \delta_\mathrm{cat}^\mathrm{poly} + \delta_\mathrm{spont}^\mathrm{poly} - \delta_\mathrm{spont}^\mathrm{depoly}) a \, \partial_x g(x,t)
 + \frac{1}{2} ( \delta_\mathrm{cat}^\mathrm{poly} + \delta_\mathrm{cat}^\mathrm{depoly} -\delta_\mathrm{spont}^\mathrm{poly} - \delta_\mathrm{spont}^\mathrm{depoly}  )  a^2 \partial_x^2  g(x,t) \nonumber \\
 && +  \omega _\tn{a} c (1 -  \rho (x,t) )  - \omega _\tn{d} \rho(x,t)\, , \label{eq:rhobulkSpontGrow}\\ 
 \partial_t g (x,t) &=& 0 =  (\delta_\mathrm{spont}^\mathrm{depoly} +\epsilon )(\avg{n_2}(t) - \avg{n_1 n_2}(t)) \rho(x,t) + \delta_\mathrm{spont}^\mathrm{depoly} (\avg{n_2}(t)-\avg{n_1 n_2}(t))( a \partial_x \rho(x,t) + \frac{1}{2} a^2 \partial_x^2 \rho(x,t)) \nonumber \\
 &&+ (\delta_\mathrm{cat}^\mathrm{depoly} -\delta_\mathrm{cat}^\mathrm{poly} ) a \partial_x g(x,t) +(\epsilon + \frac{1}{2} \delta_\mathrm{cat}^\mathrm{poly}+ \frac{1}{2}\delta_\mathrm{cat}^\mathrm{depoly}) a^2 \partial_x^2 g(x,t) + \omega_\tn{a} c (\avg{n_1}(t)- \rho(x,t) - 2 g(x,t) )  \nonumber \\
 && - (\omega_\tn{d} + \overline{\omega}_\tn{d}) g(x,t)\, , \\\label{eq:CxSpontGrow}
 \tfrac{\tn{d}}{\tn{d} t}{\avg{ {n}_1 }(t) }&=& 0 = \epsilon (\avg{n_2} (t)-\avg{n_1 n_2} (t)) +  \delta_\mathrm{spont}^\mathrm{depoly} (\avg{n_2}(t) - \avg{n_1 n_2}(t)) + \omega _\tn{a} c (1-\avg{ n_1}(t)) - \overline{\omega}_\tn{d} \avg{ n_1} (t)\, , \\   
 \tfrac{\tn{d}}{\tn{d} t}{\avg{ {n}_2}(t)} &=& 0 =\epsilon (\avg{ n_3} (t) - 2 \avg{n_2} (t) + \avg{n_1 n_2} (t)) -  \delta_\mathrm{cat}^\mathrm{poly} \avg{ n_1 n_2 } (t) + \delta_\mathrm{cat}^\mathrm{depoly}(\avg{n_1 n_3} (t)- \avg{n_1 n_2} (t))  \nonumber \\
 &&- \delta_\mathrm{spont}^\mathrm{poly}(\avg{n_2} (t)- \avg{ n_1 n_2 } (t))  + \delta_\mathrm{spont}^\mathrm{depoly} (\avg{n_1 n_2}(t) - \avg{n_1 n_3}(t) + \avg{n_3}(t)- \avg{n_2}(t))+  \omega _\tn{a} c (1- \avg{n_2 }(t)) - \omega _\tn{d} \avg{n_2} (t) \, , \nonumber \\
\\
 \tfrac{\tn{d}}{\tn{d} t}{\avg{{n}_1 {n}_2 } (t)}&=& 0 =\epsilon  (\avg{n_1 n_3}(t) - \avg{n_1 n_2}(t)) - \delta_\mathrm{cat}^\mathrm{poly} \avg{n_1 n_2} (t) + \delta_\mathrm{cat}^\mathrm{depoly} (\avg{n_1 n_3} (t)- \avg{n_1 n_2}(t)) \nonumber \\
 &&+ \delta_\mathrm{spont}^\mathrm{depoly} (\avg{n_2}(t) \avg{n_3}(t) - \avg{n_1 n_2}(t)\avg{n_3}(t)) + \omega_\tn{a} c (\avg{n_1}(t) + \avg{n_2}(t) - 2 \avg{n_1 n_2} (t)) - (\omega_\tn{d} + \overline{\omega}_\tn{d}) \avg{n_1 n_2} (t).\label{eq:n1n2SpontGrow}
\eequa
The equations are solved  in analogy to the case without spontaneous lattice dynamics. 

\begin{figure}[!ht]
\centering
\includegraphics[width = 0.7\columnwidth]{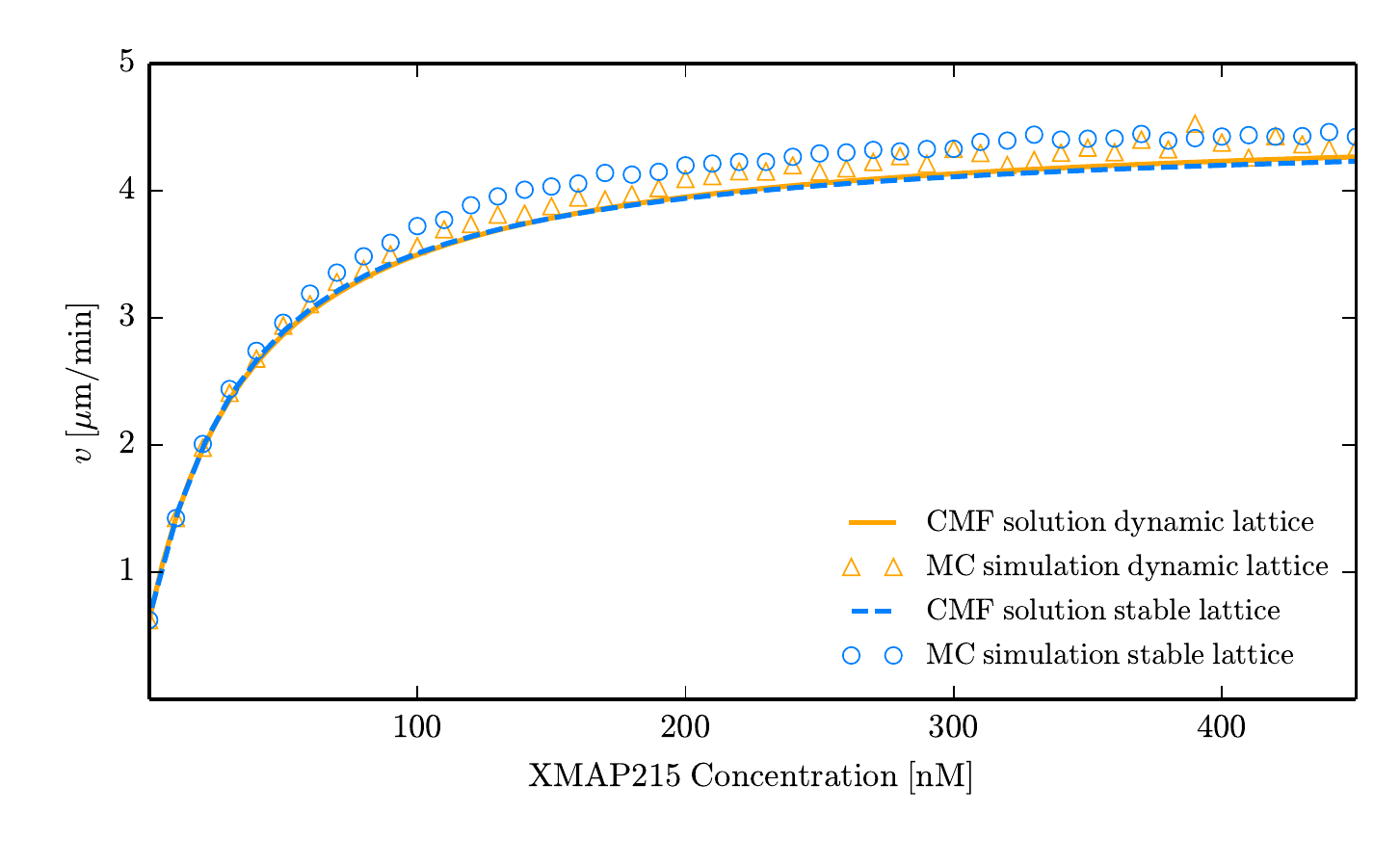}
\caption{Extended model that accounts for uncatalyzed growth and shrinkage of MTs. We compare diffusion and capture on a slowly growing lattice ($\delta_\mathrm{spont}^\mathrm{poly} = 1\, s^{-1} $, $\delta_\mathrm{spont}^\mathrm{depoly} = 0$, $\delta_\mathrm{cat}^\mathrm{depoly} = 0$,  $\delta_\mathrm{cat}^\mathrm{poly} = 9.5\, s^{-1}$, blue) with diffusion and capture on a lattice with fast intrinsic dynamics but the same average growth speed ($\delta_\mathrm{spont}^\mathrm{poly} = 51\, s^{-1} $ , $\delta_\mathrm{spont}^\mathrm{depoly} = 50\, s^{-1}$, $\delta_\mathrm{cat}^\mathrm{depoly} = 50\, s^{-1}$, $\delta_\mathrm{cat}^\mathrm{poly} = 59.5\, s^{-1}$, orange). The average MT growing velocity, and therefore also the tip density, deviate little which implies the validity of our results also on dynamic lattices. MC simulations (symbols) agree well with solutions of the CMF approximation (lines). Other parameter values are as for the XMAP215 model, see Table~\ref{tab:MCAK_rates}.\label{fig:SpontGrowth}}
\end{figure}

\begin{figure}[!ht]
\centering
\includegraphics[width = 0.7\columnwidth]{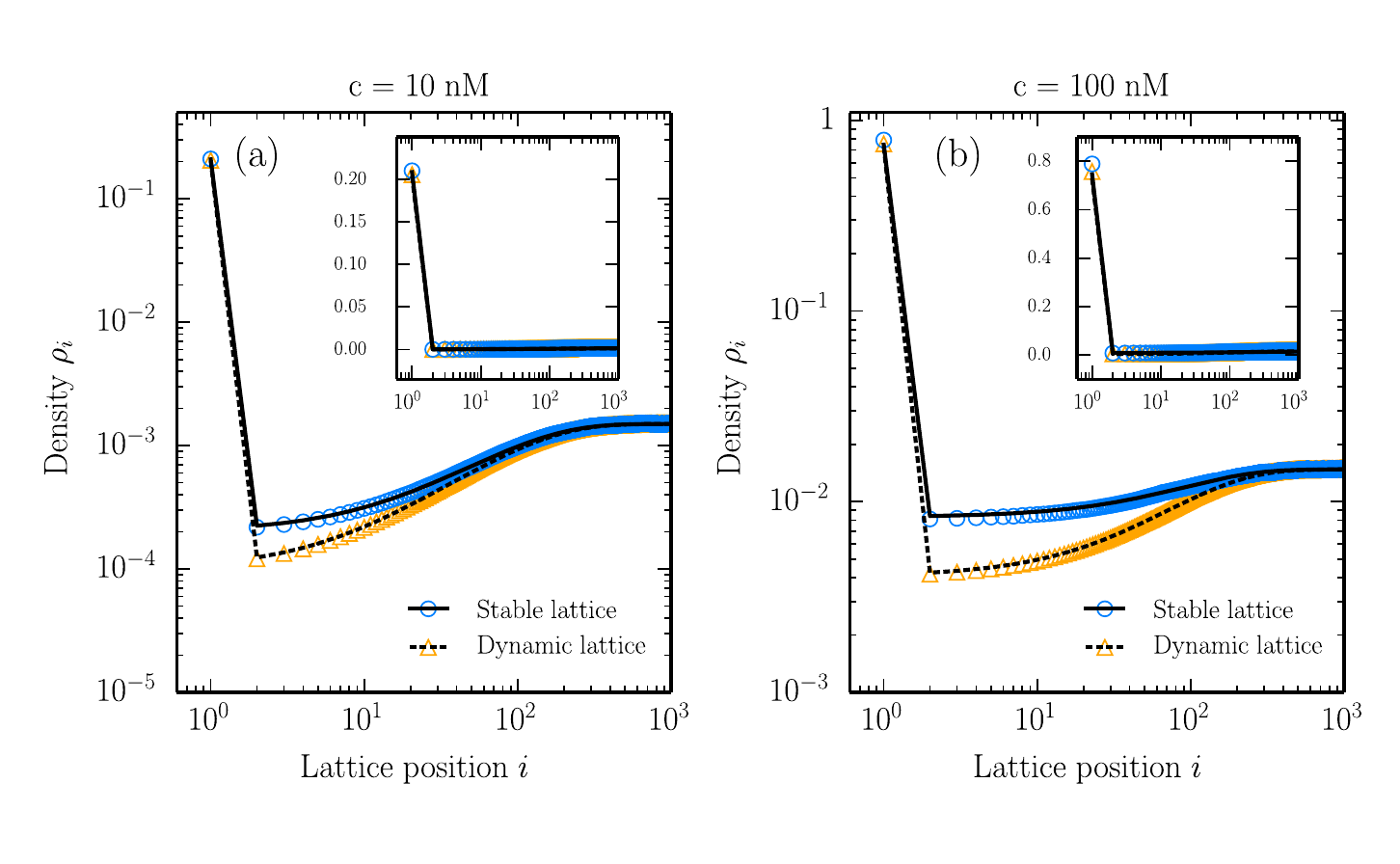}
\caption{Density profiles of an adapted model with an intrinsically dynamic lattice (orange) in comparison to the model presented in the main text (blue) for $c=10 \, \mathrm{nM}$ and  $c=100 \, \mathrm{nM}$. Tip-localization occurs also on a lattice with fast spontaneous growth and shrinkage. The tip-density is almost unaffected by rapid fluctuations of the MT length, suggesting the validity of our results also for dynamic MTs. The results of our simulations (symbols) agree well with the CMF results (lines). Model parameters are $\delta_\mathrm{spont}^\mathrm{poly} = 51\, s^{-1} $ , $\delta_\mathrm{spont}^\mathrm{depoly} = 50\, s^{-1}$, $\delta_\mathrm{cat}^\mathrm{depoly} = 50\, s^{-1}$, $\delta_\mathrm{cat}^\mathrm{poly} = 59.5\, s^{-1}$ for the dynamic lattice. Other parameters and parameters for the stable lattice as for the XMAP215 model.\label{fig:DensityDynamicLattice}}
\end{figure}

As mentioned above, our models neglect intrinsic MT dynamics such as dynamic instability. However, we expect validity of our results for tip-localization also under such circumstances. We studied the extended model with spontaneous growth and shrinkage rates over a variety of parameter values (up to spontaneous growth and shrinkage rates of $24\, \mu \mathrm{m}/ \mathrm{min}$).
For a comparison, we estimated the rate of spontaneous MT growth ($v_\mathrm{spont} = a (\delta_\mathrm{spont}^\mathrm{poly}- \delta_\mathrm{spont}^\mathrm{depoly})$) at tubulin concentrations slightly above 5 $\mu\mathrm{M}$ from the experiments performed by Widlund et al.~\cite{Widlund2011}. At such tubulin concentrations, MTs were observed to start growing also without the presence of XMAP215 at a speed of approximately $v_\mathrm{spont} =0.5\ \mu\mathrm{m}/\mathrm{min}$. Given this resulting spontaneous MT growth rate, we compared a model with and without fast intrinsic MT dynamics ($\delta_\mathrm{spont}^\mathrm{poly} = 1\, s^{-1} $ and $\delta_\mathrm{spont}^\mathrm{depoly} = 0$ for a stable lattice; $\delta_\mathrm{spont}^\mathrm{poly} = 51\, s^{-1} $ and $\delta_\mathrm{spont}^\mathrm{depoly} = 50\, s^{-1}$ for a dynamic lattice). The results are shown in Figs.~\ref{fig:SpontGrowth} and~\ref{fig:DensityDynamicLattice}. They show the robustness of the protein distribution $\rho(x)$ and, in particular, the tip occupation against changes in the lattice growth or shrinkage rates. Moreover, the CMF approximation is also applicable for rapidly fluctuating MT lengths. Note that XMAP215 also catalyzes tubulin removal under certain conditions~\cite{Brouhard2008} which could readily be accounted for in the above approach.

\section{MCAK model}
\begin{figure}[!ht]
\centering
\includegraphics[width = 0.45\columnwidth]{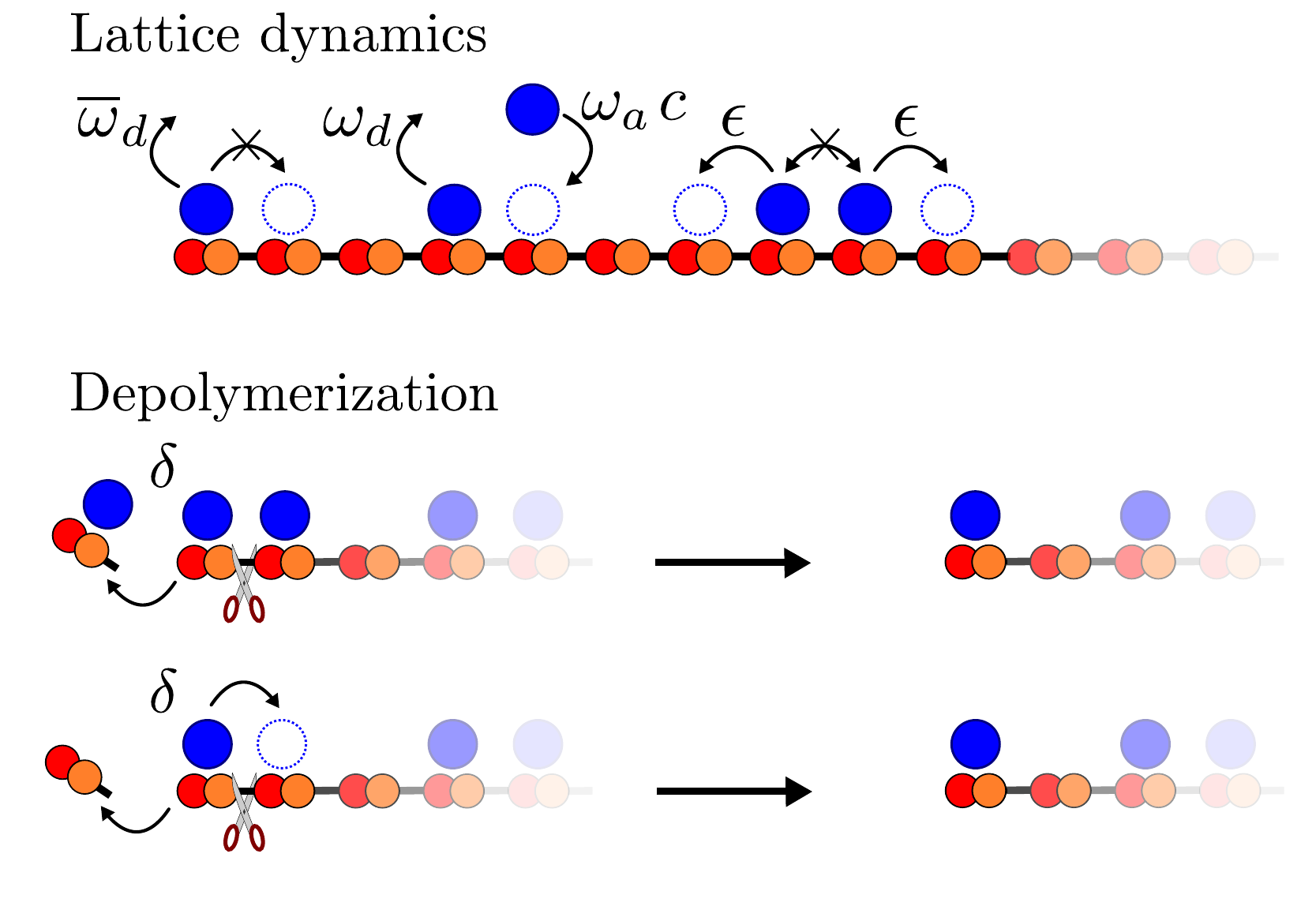}
\caption{Illustration of the MCAK model. Particle movement is identical to the XMPAP215 model. Depolymerization occurs whenever the first lattice site is occupied. Particles depolymerize processively in that they move along with the shrinking tip. When the second site is occupied, a particle on the tip that stimulates shrinkage falls off together with the first lattice site. \label{fig:MCAKmodel}}
\end{figure}
Similar to the model for XMAP215 stated in the main text we can set up a model for the depolymerase activity of MCAK, see Fig.~\ref{fig:MCAKmodel}. The ensuing set of equations corresponding to the CMF approach in the bulk are a special case of Eqs.~\ref{eq:rhobulkSpontGrow}-\ref{eq:n1n2SpontGrow} with $\delta_\mathrm{spont}^\mathrm{poly}=\delta_\mathrm{spont}^\mathrm{depoly}=\delta_\mathrm{cat}^\mathrm{poly}=0$. We implement a processive depolymerization scheme~\cite{Helenius2006,Cooper2010,Klein2005}. In detail, MCAK particles stay at the terminal site  during depolymerization (i.e. move along with the tip) whenever the neighboring site is empty. Otherwise, they dissociate from the tip during depolymerization. This means that MCAK particles fall off the MT tip whenever they hit another particle during the depolymerization process.
\begin{figure}[!ht]
\centering
\includegraphics[width = 0.7\columnwidth]{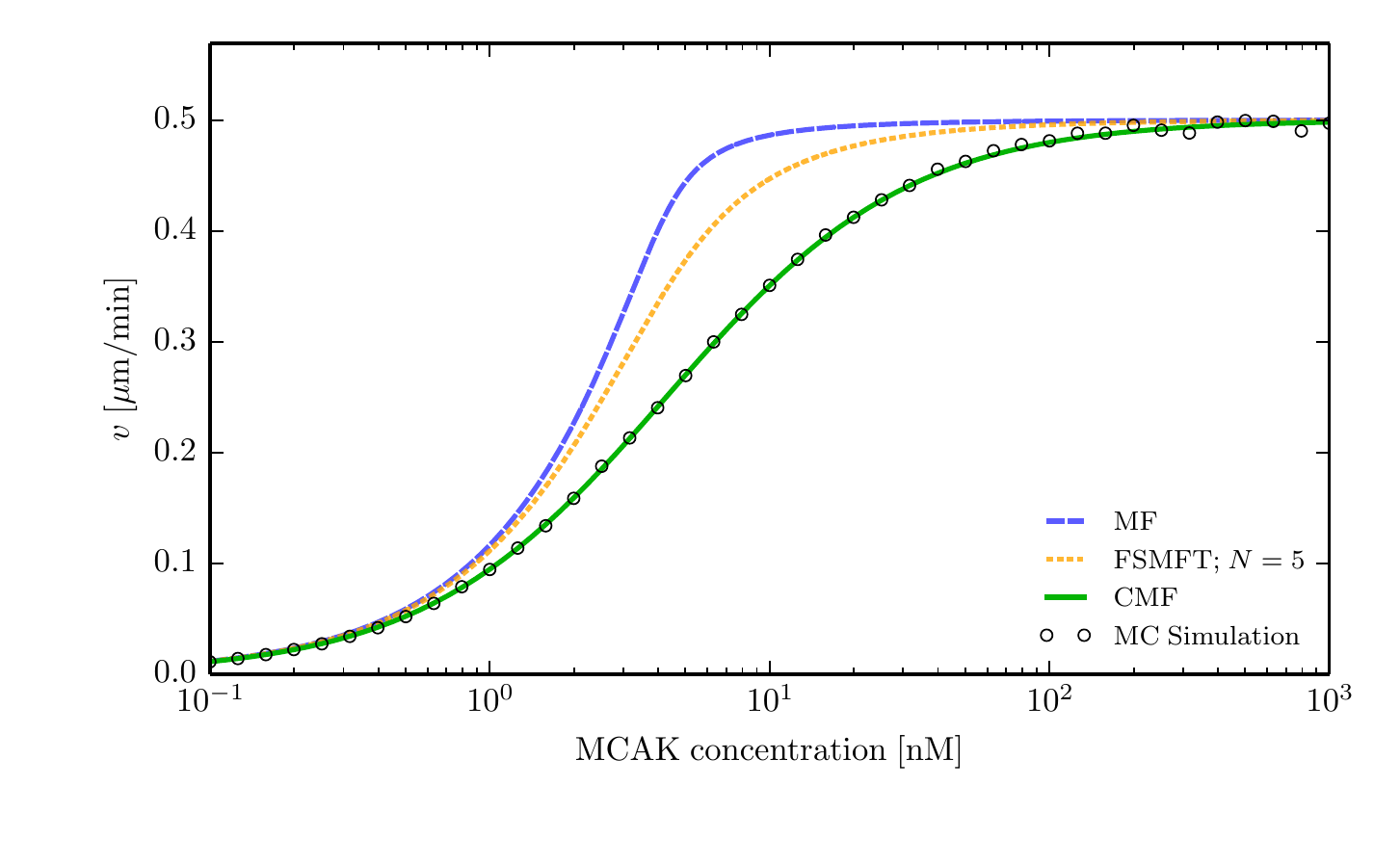}
\caption{Comparison of different analytic approaches (lines) with simulations of the MCAK model (circles). Whereas the MF and FSMFT approaches (dashed lines) predict the depolymerization velocity insufficiently, the CMF approximation (solid line) delivers results which are in excellent agreement with simulation data. Model parameters are given in Table~\ref{tab:MCAK_rates}.\label{fig:MCAK_MFSchemes}}
\end{figure}
The results of the CMF approach for the MCAK model agree excellently with simulation data, as shown in Fig.~\ref{fig:MCAK_MFSchemes}. Further, also for the MCAK model the MF approximation and FSMFT produce results that deviate from simulation data at intermediate concentrations.


\section{Parameter values}
\begin{table}
\setlength{\extrarowheight}{2.pt}
\begin{tabular*}{0.8\textwidth}{@{\extracolsep{\fill}}cccccc}
 \multirow{2}{*}{\bf{Experiment}}\\\\\hline\hline & $D$  & $k_\mathrm{on}$ & $k_\mathrm{off}$ & $v_\mathrm{max}$ & $K_M$  \\ 
 &($\mu$m)$^2\ \mathrm{\ s}^{-1}$  & events /(s $\mu$m nM) & events/s & $\mu$m/min & $\mu$m/(min nM) \\ \hline
MCAK-FL & 7.6 $\times 10^{-2}$ & 4.56 $\times 10^{-1}$ & 1.70 & 5.0 $\times 10^{-1}$ & 4.3 \\\hline
 \multirow{2}{*}{} & $D$  & $k_\mathrm{on}$ & $k_\mathrm{off}$ & $v_\mathrm{max}$ & $K_\tn{off}$  \\ 
 &($\mu$m)$^2 \ \mathrm{\ s}^{-1}$ & events /(s $\mu$m nM) & events/s & $\mu$m/min &  s$^{-1}$ \\ \hline
XMAP215 & 3.0 $\times 10^{-1}$ & 1$\times 10^{-1}$ & 4.1$\times 10^{-1}$ & 4.6 & 2.6 $\times 10^{-1}$ \\\hline\\
\multirow{2}{*}{\bf{Theory}} \\ \\ \hline\hline& $\epsilon $ & $\omega _\tn{a}$ & $\omega_\tn{d}$ & $\delta$ & $\overline{\omega}_\tn{d}$ \\ 
& $\mathrm{\ s}^{-1}$ & $\mathrm{\ (nM\ s)}^{-1}$ & $\mathrm{\ s}^{-1}$ & $\mathrm{\ s}^{-1}$ & $\mathrm{\ s}^{-1}$ \\ \hline
 MCAK-FL & 1.2 $\times 10^{3}$ & 2.61 $\times 10^{-4}$ & 1.70 & 5.2  $\times 10^{-1}$ & 3.0 $\times 10^{-2}$ \\\hline
XMAP215 & 4.7 $\times 10^{3}$\ & 6 $\times 10^{-5}$ & 4.1 $\times 10^{-1}$ & 9.5  & 2.6 $\times 10^{-1}$ \\\hline
 \end{tabular*}
  \caption{Rate constants for MCAK-FL~\cite{Cooper2010} and XMAP215~\cite{Brouhard2008,Widlund2011}. The diffusion constant $D$ and the on- and off-rates of enzymes to the MT lattice, $k_\mathrm{on}$ and $k_\mathrm{off}$, were measured directly. The measured depolymerization and polymerization profiles yield the maximal depolymerization and polymerization velocities $v_\mathrm{max}$ and the effective Michaelis constant $K_M$.
Conversion to the theoretical values was achieved by translating $k_\mathrm{on}$, $k_\mathrm{off}$, and $v_\mathrm{max}$, into appropriate lattice units. The hopping rate is related to the diffusion coefficient by $\epsilon = D/a^2$. The off-rate at the first site for MCAK was, in contrast to the one for XMAP215, not measured directly. It can, however, be estimated from $K_M$ by using the depolymerization behavior at low concentrations and a MF argument which exploits the fact that the system is uncorrelated at asymptotically low occupations~\cite{Reithmann2015}. } 
\label{tab:MCAK_rates}
\label{tab:MCAK_exp}
\end{table}

The parameter values used for the XMAP215 and MCAK model were extracted from experimental data~\cite{Cooper2010,Brouhard2008,Widlund2011}. Model parameters were computed based on measured diffusion coefficients (for $\epsilon$), particle dwell times on the MT tip (for $\overline{\omega}_d$) and bulk (for $\omega_d$), attachment rates (for $\omega_a$), and maximal (de)polymerization velocities at saturated (de)polymerase concentrations (for $\delta$). A conversion factor $n_\mathrm{tubulins}$ from $\mu\mathrm{m}$ into tubulin subunits was adapted to the assumed protofilament numbers $ n_\mathrm{protofilaments}$ of the MTs used in the respective experiments: 1625 $\mathrm{tubulin\ dimers}/\mu\mathrm{m}$ for XMAP215~\cite{Widlund2011} and 1750 $\mathrm{tubulin\ dimers}/\mu\mathrm{m}$ for MCAK~\cite{Cooper2010}. Note that the polymerization velocity refers to one MT tip~\cite{Brouhard2008,Widlund2011}, wheres the depolymerization rate refers to the average shrinkage rate of both ends~\cite{Cooper2010}. Opposed to the measurements for MCAK, where the maximal depolymerization velocity was determined~\cite{Cooper2010}, Widlund et al. do not directly state the maximal MT polymerization velocity due to XMAP215 induced growth~\cite{Widlund2011}. To get a good estimate for the maximal growing velocity $v_\mathrm{max}$ of MTs at saturating polymerase (XMAP215) concentrations, we fitted a Michaelis-Menten curve to the experimental data. The rate of tubulin attachment and detachment per regulating protein $\delta$ depends on the maximal number of catalytically active proteins at the MT tip $n_\mathrm{tip}$: $v_\mathrm{max} = \delta \, n_\mathrm{tip}\, n_\mathrm{tubulins}^{-1}$. Since the specific number for $n_\mathrm{tip}$ is elusive (there are estimates for approximately 10 XMAP215s at the MT tip at $50\ \tn{nM}$ XMAP215.~\cite{Brouhard2008}), we have to make an assumption. Here, we choose one protein per protofilament, $n_\mathrm{tip} = n_\mathrm{protofilaments}$. In doing so the MT tip velocity then reduces to  $v = \avg{n_1} \, \delta \, n_\mathrm{protofilaments} \, n_\mathrm{tubulins}^{-1} = \avg{n_1} \, \delta \, a$, where $a$ is the length of a tubulin dimer.

As the dwell time of proteins on the tip (i.e. $1/\overline{\omega}_d$) was not measured for MCAK particles, we used the measured Michaelis constant $K_M$ to estimate this value: Since the Michaelis constant determines the linear increase in the depolymerization velocity for asymptotically low MCAK concentrations, $v_{\mathrm{low\ }c}=1/K_M \times c +\mathcal{O}(c^2)$, we can use it to estimate the tip-dwell time for MCAK particles. In detail, we analytically computed the depolymeriztion velocity for asymptotically low concentrations using a MF and low-density approximation of our model up to first order in $c$~\cite{Reithmann2015}. As
correlations vanish under these conditions, we expect the result to be exact which allows us to infer the MCAK off-rate at the tip $\overline{\omega}_d$.
The list of ensuing parameters is given in Table~\ref{tab:MCAK_rates}.

\end{bibunit}

\end{document}